# A search for giant radio galaxies in recent deep radio surveys

Brissa Gómez Miller[1] and Heinz Andernach[2]
1) Universidad Autónoma de Yucatán, Mérida, gmillerbrissa@gmail.com
2) Departamento de Astronomía, DCNyE, Univ. de Guanajuato, Guanajuato, heinz@astro.ugto.mx

ABSTRACT
Giant Radio Galaxies (GRG) are those whose linear size projected on the sky exceeds one Megaparsec (1 Mpc = 3.09 x $10^{22}$ m = 3.3 million light years). Since only about 300 of these have been reported in literature, we used two recent deep radio surveys to search for further examples of these rare objects. Here we report the discovery of several new GRGs in these surveys.

INTRODUCTION

The median projected largest linear size (LLS) of classical double radio galaxies is below 200 kpc and only a small fraction of these (<1 %) exceed this size by about 5 times (1 Mpc) and are called giant radio galaxies (GRG). Their extreme sizes were initially attributed to (a) a preferred orientation in the sky plane, (b) a location in less dense environments, or (c) more powerful jets, but all these were shown to be inconsistent with observational evidence [1]. However, it became clear that due to their huge extent they may serve as tracers of the large-scale galaxy distribution and that there is marginal evidence for the ejection of the jets or lobes to occur in directions with negative gradients of galaxy density [2]. Since 2012, one of us (H.A.), with the help of various students, have used visual inspection of modern large-scale radio surveys like the *NRAO Very Large Array Sky Survey* (NVSS) [3] or *Faint Images of the Radio Sky at Twenty Centimeters* (FIRST) [4] to more than triplicate the number of known GRGs [5]. A preliminary statistical analysis revealed interesting results [6], motivating us to seek further confirmation from larger GRG samples.

MATERIALS AND METHODS

We selected two recent radio surveys that cover large areas of sky for visual inspection of their images to search for extended features which would suggest the presence of large radio galaxies. The first of these surveys [7] was made with the Jansky Very Large Array (NM, USA) in the frequency band 1-2 GHz and covers an area of 100 $deg^2$ of the *Sloan Digital Sky Survey* (SDSS) Stripe82 along the celestial equator at an angular resolution of 16" x 10", and we refer to it as HJB2016. The second one is the *LOFAR Two-meter Sky Survey Preliminary Data Release* (LoTSS-PDR) [8] and was made with the LOFAR radio telescope (Netherlands) at lower frequency (120-168 MHz) and an angular resolution of 25", covering 350 $deg^2$ between declinations +45° and +57°. For HJB2016 there were 12 images of a total of 5.7 GB to inspect, while for LoTSS-PDR there were 54 images of 28 GB in total. The images were displayed with the program `ds9` (see ds9.si.edu) together with positions of radio galaxies already known from [6] so as to avoid rediscoveries. Each candidate object was marked with a circle with diameter of the total extent of the source and saved as a "region file". The resulting source positions were cleaned from duplicates due to the large overlap between neighboring images. This yielded a total of 177 new candidates from HJB2016 and 2055 from LoTSS-PDR, of which roughly 25 % are high-, 25 % medium-, and 50 % are low-priority objects, depending on their shape and/or connection between the components of the candidates. Radio images from NVSS, FIRST, VLAS82 [9] and TGSS-ADR1 [10] as well as optical images and redshifts from SDSS [11] and photometric redshift estimates ($z_{ph}$) from various sources cited in [6] were used to assess the reality of the sources and the distance of its host galaxy. Only high-priority objects were inspected within the six weeks available for this project.

For a few objects we used the `Aladin` software [12] to measure geometrical parameters like the angular distance (or "armlength") from the host galaxy or quasar to the outer extremes of the radio lobes on both sides, as well as each armlength's position angle on the sky, and the total flux of each lobe by integrating over their extended emission regions. This allowed us to determine the armlength ratio (ALR) in the sense of stronger-to-fainter lobe length, the bending (or misalignment) angle (BA) between the lobes as the complement of the difference of the arm's position angles, as well as the ratio of the integrated fluxes (FLR) in the sense of longer-to-shorter arm flux.

RESULTS AND CONCLUSIONS

About a dozen objects larger than 0.5 Mpc were discovered in the HJB2016 survey, most notably the GRG hosted by SDSS J224520.76-003206.1, aka AllWISE J224520.79-003206.3, at redshift $z_{ph}$=0.66. With its angular size of 2.98' it has an LLS of 1.2 Mpc/$h_{70}$, where the Hubble constant is assumed as $H_0$=70 $h_{70}$ km s$^{-1}$Mpc$^{-1}$. Its ALR and FLR are 0.88 and 0.6, respectively, with a bending angle of 12º. In both [7] and [9] its outer hotspots show faint radio trails towards the host galaxy, but its radio core is barely dectected at ~0.3 mJy corresponding to a radio luminosity of log $P_{1.4GHz}$ ~ 23.5 W/Hz. In NVSS only two unconnected extended sources (the lobes) appear aligned with each other, with a total flux of ~16 mJy, implying a total log $P_{1.4}$ ~ 25.3 W/Hz.

In LoTSS-PDR several new GRGs were discovered, e.g. a 3.57'-sized radio galaxy PSO J110245.010+525757.76 (aka AllWISE J110244.92+525755.3) with a bending angle of only 2º, but an ALR of 0.69 and an FLR of 0.25. The host galaxy is undetected in SDSS and only detected in the near infrared in PanSTARRS [13] as an i=20.5 mag object, suggesting a redshift of at least ~0.5 and thus an LLS of at least 1.3 Mpc/$h_{70}$; its integrated flux at 150 MHz and 1.4 GHz yields log$P_{150 MHz}$≥26.3 W/Hz and log$P_{1.4 GHz}$≥25.3 W/Hz.

The inspection of several hundred more promising objects selected in the present project is likely to reveal on the order of a dozen or more GRGs, predominantly radio-faint and distant ones. More detailed results from the LoTSS-PDR inspection, especially an unprecedented case of a potentially aligned pair of giant radio galaxies at redshift z ~ 0.5, will be described in [14].

ACKNOWLEDGEMENTS

We thank Tim Shimwell for access to the LoTSS-PDR images. Summer students Braulio Arredondo P., Douglas Monjardín W., and Jonatan Renteria M. also contributed to this work. H.A. benefitted from grant DAIP 980/2016-2017 of University of Guanajuato.